\documentclass[onecolumn,showpacs]{revtex4}
\usepackage{graphicx}
\usepackage{dcolumn}
\usepackage{bm}

\input epsf

\begin{document}

\title{\Large  Holographic Dark Energy Scenario and Variable Modified Chaplygin Gas}

\author{\bf Surajit
Chattopadhyay$^1$\footnote{surajit$_{-}$2008@yahoo.co.in} and
~Ujjal Debnath$^2$\footnote{ujjaldebnath@yaghoo.com}}

\affiliation{ $^1$Department of Computer Application, Pailan
College of Management and Technology, Calcutta-104, India.\\
$^2$Department of Mathematics, Bengal engineering and Science
University, Howrah-103, India.}

\date{\today}

\begin{abstract}
In this letter, we have considered that the universe is filled
with normal matter and variable modified Chaplygin gas. Also we
have considered the interaction between normal matter and variable
modified Chaplygin gas in FRW universe. Then we have considered a
correspondence between the holographic dark energy density and
interacting variable modified Chaplygin gas energy density. Then
we have reconstructed the potential of the scalar field which
describes the variable modified Chaplygin cosmology.
\end{abstract}


\maketitle

The holographic principle emerged in the context of black-holes,
where it was noted that a local quantum field theory can not fully
describe the black holes [1]. Some long standing debates regarding
the time evolution of a system, where a black hole forms and then
evaporates, played the key role in the development of the
holographic principle [2,3,4]. Cosmological versions of
holographic principle have been discussed in various literatures
[e.g., 5,6,7]. Easther and (1999)[7] proposed that the holographic
principle be replaced by the generalized second law of
thermodynamics when applied to time-dependent backgrounds and
found that the proposition agreed with the cosmological
holographic principle proposed by Fischler and Susskind (Ref [5])
for an isotropic open and flat universe with a fixed equation of
state. Verlinde [8] studied the holographic principle in the
context of an ($n+1$) dimensional radiation dominated closed FRW
universe. Numerous cosmological observations have established the
accelerated expansion of the universe [9,10]. Since it has been
proven that the expansion of the universe is accelerated, the
physicists and astronomers started considering the dark energy
cosmological observations indicated that at about 2/3 of the total
energy of the universe is attributed by dark energy and 1/3 is due
to dark matter [11]. In recent times, considerable interest has
been stimulated in explaining the observed dark energy by the
holographic dark energy model [1,11,12]. An approach to the
problem of dark energy arises from the holographic principle
stated in the first paragraph. For an effective field theory in a
box size $L$ with UV cutoff $\Lambda_{c}$, the entropy
$L^{3}\Lambda_{c}^{3}$. The non-extensive scaling postulated by
Bekenstein suggested that quantum theory breaks down in large
volume [11]. To reconcile this breakdown, Chohen et al [13]
pointed out that in quantum field theory a short distance (UV)
cut-off is related to a long distance (IR) cut-off due to the
limit set by forming a black hole. Taking the whole universe into
account the largest IR cut-off $L$ is chosen by saturating the
inequality so that we get the holographic dark energy density as
[11] $\rho_{\Lambda}=3c^{2}M_{p}^{2}L^{-2}$ where $c$ is a
numerical constant and $M_{p}\equiv1/\sqrt{8\pi G}$ is the reduced
Plank mass. On the basis of the holographic principle proposed by
[5] several others have studied holographic model for dark energy
[10]. Employment of Friedman equation [14] $\rho=3M_{p}^{2}H^{2}$
where $\rho$ is the total energy density and taking $L=H^{-1}$ one
can find $\rho_{m}=3(1-c^{2})M_{p}^{2}H^{2}$. Thus either
$\rho_{m}$ or $\rho_{\Lambda}$ behaves like $H^{2}$. Thus, dark
energy results as pressureless. But, neither dark energy, nor dark
matter has laboratory evidence for its existence directly. Thus,
Cardone et al [15] and Bento et al [16] proposed unified dark
matter $/$ energy scenario in which two dark components are
different manifestations of a single cosmic fluid. Some
interesting examples of such an unification are the generalized
Chaplygin gas, the tachyonic field, and the condensate cosmology
[15].\\

The Chaplygin gas is characterized by an exotic equation of state
$p=-\frac{B}{\rho}$ [17], where $B$ is a positive constant. Role
of Chaplygin gas in the accelerated universe has been studied by
several authors. The above mentioned equation of state has been
modified to $p=-\frac{B}{\rho^{\alpha}}$ with $0\leq\alpha\leq 1$.
This is called generalized Chaplygin gas [18]. This equation has
been further modified to $p=A\rho-\frac{B}{\rho^{\alpha}}$ with
$0\leq\alpha\leq 1$. This is called modified Chaplygin gas [19].
This equation of state shows radiation era at one extreme and
$\Lambda CDM$ model at the other extreme. Correspondence between
the holographic dark energy scenario and the Chaplygin gas is
studied in [14,20]. Debnath [21] introduced a variable modified
Chaplygin gas with $B$ as a function of the scale factor $a$.
Thus, the equation of state is
$p=A\rho-\frac{B(a)}{\rho^{\alpha}}$. Present paper endeavors to
establish a correspondence between the holographic dark energy
scenario and the variable modified
Chaplygin gas model.\\

The metric of a homogeneous and isotropic universe in an FRW model
is

\begin{equation}
ds^{2}=dt^{2}-a^{2}(t)\left[\frac{dr^{2}}{1-kr^{2}}+r^{2}(d\theta^{2}+\sin^{2}\theta
d\phi^{2})\right]
\end{equation}

where, $a(t)$ is the scale factor, and $k$ denotes the curvature
of the space. \\

The first Friedman equation is given by

\begin{equation}
H^{2}+\frac{k}{a^{2}}=\frac{1}{3M_{p}^{2}}[\rho_{\Lambda}+\rho_{m}]
\end{equation}

Let us define
$\Omega_{m}=\frac{\rho_{m}}{3M_{p}^{2}H^{2}}$,
$\Omega_{\Lambda}=\frac{\rho_{\Lambda}}{3M_{p}^{2}H^{2}}$,
$\Omega_{k}=\frac{k}{a^{2}H^{2}}$.

Assuming $B(a)=B_{0}a^{-n}$ in the equation of state of the
variable modified Chaplygin gas with $B_{0}>0$ and $n(>0)$ as
constant we get the solution $\rho$ as

\begin{equation}
\rho_{\Lambda}=\left[\frac{3(1+\alpha)B_{0}}{3(1+\alpha)(1+A)-n}\frac{1}{a^{n}}+
\frac{C_{0}}{a^{3(1+A)(1+\alpha)}}\right]^{\frac{1}{1+\alpha}}
\end{equation}

The continuity equations for variable modified Chaplygin gas and
cold dark matter are

\begin{equation}
\dot{\rho}_{\Lambda}+3H(1+\omega_{\Lambda})\rho_{\Lambda}=-\delta\rho_{\Lambda}
\end{equation}

\begin{equation}
\dot{\rho}_{m}+3H(1+\omega_{m})\rho_{m}=\delta\rho_{\Lambda}
\end{equation}

where $\delta$ is the interaction term. This corresponds to the
decay of generalized Chaplygin gas component into CDM. Taking
$x=\frac{\rho_{m}}{\rho_{\Lambda}}$ it is obtained from equation
(5) that

\begin{equation}
\dot{x}=3Hx\left[(\omega_{\Lambda}-\omega_{m})+\frac{1+x}{x}\frac{\delta}{3H}\right]
\end{equation}

 We define

\begin{equation}
w_{\Lambda}^{eff}=w_{\Lambda}+\frac{\delta}{3H}
\end{equation}

\begin{equation}
w_{m}^{eff}=-w_{m}-\frac{1}{x}\frac{\delta}{3H}
\end{equation}

Taking derivative of both sides of equation (3) with respect to
cosmic time we obtain

\begin{equation}
\dot{\rho}_{\Lambda}=3H\left(C_{0}a^{-(1+A)(1+\alpha)}+\frac{3(1+\alpha)a^{-n}B_{0}}{3(1+A)(1+\alpha)-n}\right)^{-\frac{\alpha}{1+\alpha}}
\left(-(1+A)C_{0}a^{-3(1+A)(1+\alpha)}-\frac{na^{-n}B_{0}}{3(1+A)(1+\alpha)-n}\right)
\end{equation}

In  non-flat universe, our choice for holographic dark energy
density is

\begin{equation}
\rho_{\Lambda}=3c^{2}M_{p}^{2}L^{-2}
\end{equation}

Using holographic dark energy density we obtain
\begin{equation}
\dot{\rho}_{\Lambda}=-3^{1-\alpha}H\left((1+A)C_{0}a^{-3(1+\alpha)(1+A)}+\frac{nB_{0}a^{-n}}{3(1+A)(1+\alpha)-n}\right)
\left(c^{2}L^{-2}M_{p}^{2}\right)^{-\alpha}
\end{equation}
and
\begin{equation}
w_{\Lambda}=A-\frac{B_{0}a^{-n}}{(3c^{2}M_{p}^{2}L^{-2})^{1+\alpha}}
\end{equation}

Using the definition
$\Omega_{\Lambda}=\frac{\rho_{\Lambda}}{3M_{p}^{2}H^{2}}$ it can
be obtained that

\begin{equation}
HL=\frac{c}{\sqrt{\Omega_{\Lambda}}}
\end{equation}

Here, ${\it c}$ is a positive constant in holographic model of
dark energy $c\geq 1$. {\it L} is defined as

\begin{equation}
L=aR(t)
\end{equation}

where $a$ is the scale factor and $R(t)$ is relevant to the future
event horizon of the universe [14] and it can be derived that

\begin{equation}
L=\frac{a(t)sinn[\sqrt{|k|}R_{h}(t)/a(t)]}{\sqrt{|k|}}
\end{equation}

where, $R_{h}$ is the event horizon.\\

Now, equation (11) becomes

\begin{equation}
\dot{\rho}_{\Lambda}=-3^{1-\alpha}\frac{c}{L\sqrt{\Omega_{\Lambda}}}[(1+A)C_{0}a^{-3(1+A)(1+\alpha)}+
\frac{nB_{0}a^{-n}}{3(1+A)(1+\alpha)-n}](c^{2}L^{-2}M_{p}^{2})^{-\alpha}
\end{equation}

Substituting this relation into the interaction equation (4) and
taking $\delta=3b^{2}H(\frac{1+\Omega_{k}}{\Omega_{\Lambda}})$ we
get

\begin{equation}
w_{\Lambda}=-1-\frac{b^{2}(1+\Omega_{k})}{\Omega_{\Lambda}}+\left(C_{0}(1+A)-\frac{na^{-n+3(1+\alpha)
(1+A)}B_{0}}{n-3(1+\alpha)(1+A)}\right)
(3\Omega_{\Lambda}M_{p}^{2}H^{2}a^{3(1+A)})^{-1-\alpha}
\end{equation}

Where, $B_{0}$ comes out to be

\begin{equation}
B_{0}=\frac{3(1+\alpha)(1+A)-n}{3(1+\alpha)}\left((3c^{2}M_{p}^{2}L^{-2})^{1+\alpha}-\frac{C_{0}}{a^{3(1+\alpha)(1+A)}}\right)
a^{n}
\end{equation}

where,

\begin{equation}
C_{0}=-\frac{(3H^{2}M_{p}^{2}\Omega_{\Lambda}a^{3(1+A)})^{1+\alpha}}{c(3(1+A)(1+\alpha)-n)}\left(3b^{2}c(1+\alpha)
(\frac{1+\Omega_{k}}{\Omega_{\Lambda}})+c(n-2(1+\alpha))
+2(1+\alpha)\sqrt{\Omega_{\Lambda}-c^{2}\Omega_{k}}\right)
\end{equation}

Using $C_{0}$ in $B_{0}$ we finally obtain

\begin{equation}
B_{0}=a^{n}(3M_{p}^{2}H^{2}\Omega_{\Lambda})^{1+\alpha}\left(\frac{1+3A}{3}+\frac{b^{2}(1+\Omega_{k})}{\Omega_{\Lambda}}
+\frac{2\sqrt{\Omega_{\Lambda}-c^{2}\Omega_{k}}}{3c}\right)
\end{equation}

Now we can rewrite the scalar potential and kinetic energy terms
as follows

\begin{equation}
V(\phi)=2H^{2}M_{p}^{2}\Omega_{\Lambda}\left[1-3A-\frac{3b^{2}(1+\omega_{k})}{\omega_{\Lambda}}+
\frac{2\sqrt{\omega_{\Lambda}-c^{2}\omega_{k}}}{c}\right]
\end{equation}
and
\begin{equation}
\dot{\phi}=\frac{cM_{p}}{L}\sqrt{2+\frac{3b^{2}(1+\omega_{k})}{2\omega_{\Lambda}}+\frac{\sqrt{\omega_{\Lambda}-c^{2}\omega_{k}}}{c}}
\end{equation}

Since the models trying to provide a description of the cosmic
acceleration are proliferating, there exists the problem of
discriminating between the various contenders.\\

In this letter, we have considered that the universe is filled
with normal matter and variable modified Chaplygin gas. Also we
have considered the interaction between normal matter and variable
modified Chaplygin gas in FRW universe. Then we have considered a
correspondence between the holographic dark energy density and
interacting variable modified Chaplygin gas energy density. From
this, we have found the expressions of the arbitrary positive
constants $B_{0}$ and $C_{0}$ of variable modified Chaplygin gas.
We have seen that if we put $n=0$ and $A=0$, variable modified
Chaplygin gas reduces to generalized Chaplygin gas and in this
case equations (19) and (20) reduce to equations (31) and (32) of
reference [22]. Then we have reconstructed the potential of the
scalar field which describes the variable modified Chaplygin cosmology. \\

{\bf Acknowledgement:}\\

The authors are thankful to IUCAA, India for warm hospitality
where part of the work was carried out. Also UD is thankful to
UGC, Govt. of India for providing research project grant (No. 32-157/2006(SR)).\\

{\bf References:}\\
\\
$[1]$ K. Enqvist, S. Hannested and M. S. Sloth, {\it JCAP} {\bf 2} 004 (2005).\\
$[2]$ L. Thorlocius, {\it hep-th}/0404098.\\
$[3]$ G. T. Hooft, {\it gr-qc}/9310026.\\
$[4]$ L. Susskind, {\it J. Math. Phys.} {\bf 36} 6377 (1995).\\
$[5]$ W. Fischler and L. Susskind, {\it hep-th}/9806039.\\
$[6]$ R. Tavakol and G. Ellis, {\it Phys. Lett. B} {\bf 469} 33 (1999).\\
$[7]$ R. Esther and D. Lowe, {\it Phys. Rev. Lett.} {\bf 82} 4967 (1999).\\
$[8]$ E. Verlinde, {\it hep-th}/0008140.\\
$[9]$ B. Wang, Y. Gong and E. Abdalla, {\it Phys. Lett. B} {\bf 624} 141 (2005).\\
$[10]$ Y. Gong, {\it Phys. Rev. D} {\bf 70} 064029 (2004).\\
$[11]$ X. Zhang, {\it Int. J. Mod. Phys. D} {\bf 14} 1597 (2005).\\
$[12]$ D. Pavon and W. Zimdahl, {\it hep-th}/0511053.\\
$[13]$ A. G. Kohen et al., {\it Phys. Rev. Lett.} {\bf 82} 4971
(1999).\\
$[14]$ M. R. Setare, {\it Phys. Lett. B} {\bf 648} 329 (2007).\\
$[15]$ V. F. Cardone, A. Troisi and S. Copoziello, {\it Phys. Rev.
D} {\bf 69} 083517 (2004).\\
$[16]$ M. C. Bento, O. Bertolami and A. A. Sen, {\it Phys. Rev. D} {\bf 70} 083519 (2004).\\
$[17]$ A. Kamenshchik, U. Moschella and V. Pasquier, {\it Phys.
Lett. B} {\bf 511} 265 (2001); V. Gorini, A. Kamenshchik, U.
Moschella and V. Pasquier, {\it gr-qc}/0403062.\\
$[18]$ V. Gorini, A. Kamenshchik and U. Moschella, {\it Phys. Rev.
D} {\bf 67} 063509 (2003); U. Alam, V. Sahni , T. D. Saini and
A.A. Starobinsky, {\it Mon. Not. Roy. Astron. Soc.} {\bf 344},
1057 (2003).\\
$[19]$ H. B. Benaoum, {\it hep-th}/0205140; U. Debnath, A.
Banerjee and S. Chakraborty, {\it Class.
Quantum Grav.} {\bf 21} 5609 (2004).\\
$[20]$ M. R. Setare, {\it Europhys. J. C} {\bf 52} 689 (2007).\\
$[21]$ U. Debnath, {\it Astrophys. Space Sci.} {\bf 312} 295
(2007).\\
$[22]$ M. R. Setare, {\it Phys. Lett. B} {\bf 654} 1 (2007).\\

\end{document}